\documentstyle[amsfonts,12pt]{article}

\title{Existence of Non-Abelian Vortices
in  the Aharony--Bergman--Jafferis--Maldacena Theory}
\date{}

\newtheorem{theorem}{Theorem}[section]
\newtheorem{oldtheorem}{Theorem}[section]
\newtheorem{oldassertion}[oldtheorem]{Assertion}
\newtheorem{oldproposition}[oldtheorem]{Proposition}

\newtheorem{oldlemma}[oldtheorem]{Lemma}
\newtheorem{olddefinition}[oldtheorem]{Definition}
\newtheorem{oldclaim}[oldtheorem]{Claim}
\newtheorem{oldcorollary}[oldtheorem]{Corollary}

\newenvironment{lemma}{\begin{oldlemma}$\!\!\!${\bf.}}{\end{oldlemma}}

\newbox\qedbox
\newenvironment{proof}{\smallskip\noindent{\bf Proof.}\hskip \labelsep}%
                        {\hfill\penalty10000\copy\qedbox\par\medskip}

\newcommand{\dd}{\mbox{d}}
\newcommand{\ee}{\end{equation}}
\newcommand{\be}{\begin{equation}}
\newcommand{\bea}{\begin{eqnarray}}
\newcommand{\eea}{\end{eqnarray}}

\newcommand{\nn}{\nonumber}
\begin{document}

\maketitle

\begin{center}
~Shouxin Chen$^{a,b}$
~Ruifeng Zhang$^{a,b, }$
and~Meili Zhu$^a$

{\small\it$^ a$School of Mathematics, Henan University, Kaifeng, Henan 475004, People's Republic of China}

{\small\it$^ b$Institute of Contemporary Mathematics, Henan University, Kaifeng, Henan 475004, People's Republic of China}

\end{center}
\begin{abstract}
Vortices in non-Abelian gauge field theory play important roles in confinement mechanism and are governed by systems of nonlinear elliptic equations of complicated structures. In this paper, we present a series of existence and uniqueness theorems for multiple vortex solutions of the non-Abelian BPS vortex equations over ${\mathbb{R}}^2$ and on a doubly periodic domain. Our methods are based on calculus of variations and a fixed-point argument. The necessary and sufficient conditions for the existence of a unique solution in the doubly periodic situation are explicitly expressed in terms of several physical parameters involved.

\end{abstract}

\section {Introduction}
\setcounter{equation}{0}\hspace*{1cm}

Vortices have important applications in many fundamental areas of physics. For example, in particle physics, vortices allow one to generate dually (electrically and magnetically) charged vortex-like solitons \cite{KK} \cite{PK} \cite{VS} known as dyons \cite{SC} \cite{ZW1} \cite{ZW2}; in cosmology, vortices generate topological defects know as cosmic strings \cite{HK} \cite{OR} which give rise to useful mechanisms  for matter formation in the early universe. Besides, both electrically and magnetically charged vortices arise in a wide range of areas in condensed-matter physics including high-temperature superconductivity \cite{FI} \cite{MSEB}, optics \cite{DKT} \cite{NAOK} \cite{PE}, and so on.\\

 Mathematically, Chern--Simons theories in (2+1)-dimensions are introduced to accommodate electricity. The equations of motions of various Chern--Simons vortex moldels are hard to approach even in the radially symmetric static cases. However, since the discovery of the self-dual structure in the Abelian Chern--Simons vortex model \cite{JLW} \cite{JW} in 1990, there came a burst of fruitful works on Chern--Simons vortex equations, non-relativistic and relativistic, Abelian and non-Abelian \cite{DU1} \cite{DU2}. For example, Aldrovandi and Schaposnik  \cite{AS} \cite{LMMS} found the non-Abelian vortex solutions when gauge field dynamics is solely governed by a  Chern--Simons action and the symmetry breaking potential is six-order in order to ensure self-duality and supersymmetric extension, in the presence of a set of orientational collective coordinates. Furthermore, the existence of Chern--Simons--Higgs vortex solutions was proved in (2+1)-dimensions  with internal collective coordinates \cite{LRT}. The existence of topological solutions for relativistic non-Abelian  Chern--Simons equations involving two Higgs particles and two gauge fields was proved through studying the full ${\mathbb{R}}^2$ limit of a coupled system of two nonlinear elliptic equations  \cite{LPY}. In 2008, Aharony, Bergman, Jafferis and Maldacena  developed the so-called ABJM theory \cite{ABJM} in terms of three dimensional  Chern--Simons-matter theories with gauge groups $U(N)\times U(N)$ and $SU(N)\times SU(N)$ which have explicit $\mathcal {N} $=6 superconformal symmetry. Before long, Auzzi and Kumar \cite{AK} find half-BPS vortex solitons, at both weak and strong couplings, in this theory. \\

 More recently, the existence of solutions for Abelian  Chern--Simons equations involving two Higgs particles and two gauge fields on a torus was proved  by Lin and Prajapat \cite{LP}.  Using the methods of monotone iterations, a priori estimates, degree-theory argument and constrained minimization, multiple vortex equations in $U(N)$ and $SO(2N)$ theories were discussed \cite{GJK} \cite{LY1} \cite{LY2} and  a series of sharp existence and uniqueness theorems were established. Lieb and Yang  \cite{LY} discussed non-Abelian vortices in supersymmetric gauge field theory, over doubly periodic domains, via a highly efficient direct minimization approach. These  studies unveil a broad spectrum of systems of elliptic equations with exponential nonlinearities and rich properties and structures, which present new challenges.\\

In this paper we will concentrate on the non-Abelian BPS vortex equations derived by Auzzi and Kumar \cite{AK} in a supersymmetric Chern--Simons--Higgs theory formulated by Aharony, Bergman, Jafferis and Maldacena \cite{ABJM}, known as the ABJM model. In terms of the methods of  \cite{LY} \cite{LY1} \cite{LY2} \cite{YA}, we obtain the existence and uniqueness of a multiple vortex solution. Meanwhile, it is hopeful that our method may be explored further to study various multiple vortex equations, arising in non-Abelian gauge field theory, of more diffcult structures. \\

The content of the rest of paper is outlined as follows. In Section
3, we prove the existence and uniqueness of a multiple vortex
solution realizing an arbitrarily prescribed vortex distribution
over ${\mathbb{R}}^2$, applying the variational method of  Jaffe and
Taubes \cite{JT} used for the Abelian Higgs model. In Section 4, we
prove the existence of a multiple vortex solution over a doubly
periodic domain under a necessary and sufficient condition
explicitly stated  in terms of some physical coupling parameters, by
a multi-constrained variational approach.  In Section 5, we prove
the existence  of multiple vortex solutions in a doubly periodic
domain by a fixed-point method, where we apply the technique of the
maximum principle and the Poincar$\acute{e}$ inequality.
Furthermore, in Section 6, our methods are shown to be equally
effective in treating the existence and uniqueness problems for the
multiple vortex solution induced from independently prescribed
distributions of zeros of two complex scalar fields, instead of one.

\section {Non-Abelian vortex equations}
\setcounter{equation}{0}\hspace*{1cm}

Recently developed by Aharony, Bergman, Jafferis and Maldacena, known also as the ABJM model \cite{ABJM}, is a Chern--Simons theory within which the matter fields are four complex scalars,
\be\label{201}
C^I=(Q^1,Q^2,R^1,R^2),\quad I=1,2,3,4,
\ee
in the bifundamental matter field ($\mathbf{N},\mathbf{\overline{N}}$) representation of the gauge group $U(N)\times U(N)$, which hosts two gauge fields, $A_\mu$ and $B_\mu$. The Chern--Simons action associated to the two gauge group $A_\mu$ and $B_\mu$ of levels $+k$ and $-k$ is given by the Lagrangian density
\be\label{202}
{\mathcal {L}}_{CS} =\frac{k}{4\pi}\epsilon^{\mu\nu\gamma}{\mathrm{Tr}}\biggl(A_\mu\partial_\nu A_\gamma+\frac{2i}{3}A_\mu A_\nu A_\gamma-B_\mu\partial_\nu B_\gamma-\frac{2i}{3}B_\mu B_\nu B_\gamma\biggr),
\ee
where the gauge-covariant derivatives on the bifundamental fields are defined as
\be\label{203}
D_\mu C^I=\partial_\mu C^I+iA_\mu C^I-iC^IB_\mu, \quad I=1,2,3,4.
\ee
The scalar potential of the mass deformed theory can be written in a compact way as \cite{GR}
\be\label{204}
V={\mathrm{Tr}}(M^{\alpha \dag}M^\alpha+N^{\alpha \dag}N^\alpha),
\ee
where
\bea\label{205}
M^\alpha&=&\rho Q^\alpha+\frac{2\pi}{k}(2Q^{[\alpha} Q^\dag_\beta Q^{\beta]}+R^\beta R^\dag_\beta Q^\alpha-Q^\alpha R^\dag_\beta R^\beta\nn\\
&&+2Q^\beta R^\dag_\beta R^\alpha-2R^\alpha R^\dag_\beta Q^\beta),\\\label{206}
N^\alpha&=&-\rho R^\alpha+\frac{2\pi}{k}(2R^{[\alpha }R^\dag_\beta R^{\beta]}+Q^\beta Q^\dag_\beta R^\alpha-R^\alpha Q^\dag_\beta Q^\beta\nn\\
&&+2R^\beta Q^\dag_\beta Q^\alpha-2Q^\alpha Q^\dag_\beta R^\beta),
\eea
where the Kronecker symbol $\epsilon^{\alpha\beta} ~(\alpha, \beta =1,2)$ is used to lower or raise indices, and $\rho >0$ a massive parameter. Thus, when the spacetime metric is of the signature $(+ - -)$, the total (bosonic) Lagrangian density of ABJM model can be written as
\be\label{207}
{\mathcal {L}}=-{\mathcal {L}}_{CS}+{\mathrm{Tr}}([D_\mu C^I]^\dag[D^\mu C^I])-V,
\ee
which is of a pure Chern--Simons type for the gauge field sector. As in \cite{AK}, we focus on a reduced situation where (say) $R^\alpha=0$. Then, by virture of (\ref{205}) and (\ref{206}), the scalar potential density (\ref{204}) takes the form
\be\label{208}
V={\mathrm{Tr}}(M^{\alpha\dag}M^\alpha),\quad M^\alpha=\rho Q^\alpha+\frac{4\pi}{k}(Q^\alpha Q^\dag_\beta Q^\beta-Q^\beta Q^\dag_\beta Q^\alpha).
\ee
The equations of motion of the Lagrangian (\ref{207}) are rather complicated. However, in the static limit, Auzzi and Kumar \cite{AK} showed that these equations may be reduced into the following first-order ${\mathrm{BPS}}$ system of equations
\bea\label{209}
D_0Q^1-iW^1&=&0,\quad D_1Q^2-i D_2Q^2=0,\\\label{2010}
D_1Q^1&=&0, \quad D_2Q^1=0,\quad D_0Q^2=0,\quad W^2=0,
\eea
coupled with the Gauss law constraints which are the temporal components of the Chern--Simons equations
\bea\label{2011}
\frac{k}{4\pi} \epsilon^{\mu \nu\gamma}F^{(A)}_{\nu\gamma}=i(Q^\alpha[D^\mu Q^\alpha]^\dag-[D^\mu Q^\alpha]Q^{\alpha\dag}),\\\label{2012}
\frac{k}{4\pi} \epsilon^{\mu \nu\gamma}F^{(B)}_{\nu\gamma}=i([D^\mu Q^\alpha]^\dag Q^\alpha-Q^{\alpha\dag}[D^\mu Q^\alpha]),
\eea
where
\bea
F^{(A)}_{\mu\nu}&=&\partial_\mu A_\nu-\partial_\nu A_\mu+i[A_\mu,A_\nu],\nn\\
F^{(B)}_{\mu\nu}&=&\partial_\mu B_\nu-\partial_\nu B_\mu+i[B_\mu,B_\nu],\nn\\
W^1&=&\rho Q^1+\frac{2\pi}{k}(Q^1Q^{2\dag}Q^2-Q^2Q^{2\dag}Q^1),\nn\\
W^2&=&\rho Q^2+\frac{2\pi}{k}(Q^2Q^{1\dag}Q^1-Q^1Q^{1\dag}Q^2),\nn
\eea provided that \cite{AK} one takes that $Q^1$ assumes its vacuum
expectation value \be\label{2013} Q^1=\sqrt{\frac{\rho
k}{2\pi}}\textmd{diag}\biggl(0,1,\cdots,\sqrt{N-2},\sqrt{N-1}\biggr),
\ee the non-trivial entries of $Q^2$ are given by $(N-1)$ complex
scalar fields $\kappa$ and $\phi_\ell ~(\ell=1,\cdots,N-2)$
according to \be\label{2014} Q^2_{N,N-1}=\sqrt{\frac{\rho
k}{2\pi}}\kappa,\quad Q^2_{N-\ell,N-\ell-1}=\sqrt{\frac{\rho
k}{2\pi}}\phi_\ell,
\ee and the spatial components of the gauge
fields $A_j$ and $B_j ~(j=1,2)$ are expressed in terms of $(N-1)$
real-valued vector potentials $a^\ell=(a^\ell_j)$ and
$b=(b_j)~(j=1,2;\ell=1,\cdots,N-2)$ satisfying \be\label{2015}
A_j=B_j=\textmd{diag}(0,a^{N-2}_j,\cdots,a^1_j,b_j),\quad j=1,2. \ee

We now consider the solution for the $N=3$ case. The ansatz for the bifundamental scalars approaching the Higgs vacuum at infinity is
\begin{eqnarray}
Q^1&=&\sqrt{\frac{\rho k}{2\pi}}\left(\begin{array}{ccc}0&0&0\\
0&1&0\\0&0&\sqrt{2}\end{array}\right),\nn \\
Q^2&=&\sqrt{\frac{\rho k}{2\pi}}\left(\begin{array}{ccc}0&0&0\\
\sqrt{2}\kappa&0&0\\0&\phi&0\end{array}\right),\\
A_j&=&B_j=\left(\begin{array}{ccc}0&0&0\\
0&a_j&0\\0&0&b_j\end{array}\right), \quad j=1,2.\nn
\end{eqnarray}
Where $\kappa$ is a real-valued scalar field, $\phi$ a
complex-valued scalar field, and $a_j$ and $ b_j$ are two
real-valued gauge potential vector fields.

Define $a_{jk}=\partial_ja_k-\partial_ka_j$ and set
$\lambda=4\rho^2$. Then the vortex equations without assuming radial
symmetry are
\bea\label{o2}
  (\partial_1+i\partial_2)\kappa &=& i(a_1+ia_2)\kappa,\\
   \label{o3}
   (\partial_1+i\partial_2)\phi &=& -i([a_1+ia_2]-[b_1+i
   b_2])\phi.\\
   \label{o4}  a_{12} &=& -\frac{\lambda}{2} (2\kappa^2-|\phi|^2-1),\\
   \label{o5}  b_{12}& =& -\lambda(|\phi|^2-1).
\eea

We shall look for solutions of these equations so that $\kappa$
never vanishes but $\phi$ vanishes exactly at the finite set of
points
\be\label{o6}
  Z=\{p_1, p_2,\cdots, p_n\}.
\ee
A solution is called an $n$-vortex solution \cite{JT}.

To facilitate our computation, it will be convenient to adopt the
complexified derivatives \be\label{o7}
  \partial=\frac{1}{2}(\partial_1-i\partial_2),\quad\overline{\partial}=\frac{1}{2}(\partial_1+i\partial_2),
\ee
 and the notation
\be\label{o8}
a=a_1+ia_2,\quad b=b_1+ib_2.
\ee
As a consequence, away from $Z$, the equations (\ref{o2}) and
(\ref{o3}) become
\be\label{o9}
\overline{\partial}\ln\kappa=-\frac{i}{2}
    a,\quad\overline{\partial}\ln\phi=-\frac{i}{2}(a-b),
\ee
which allow us to solve for $a, b$ to get
\be\label{o10}
   a=2i\overline{\partial}\ln\kappa,\quad
   a-b=2i\overline{\partial}\ln\phi.
\ee
Using
\be\label{o11}
  a_{12}=-i(\partial a-\overline{\partial}\overline{a}),
\ee
(\ref{o4}), (\ref{o5}), (\ref{o10}), and the fact that
$\partial\overline{\partial}=\overline{\partial}\partial=\frac{1}{4}\Delta$,
we have
\be\label{o12}
  a_{12}=-\Delta \ln \kappa.
\ee
Likewise, we have, away from $Z$, the relation
\be\label{o13}
b_{12}=a_{12}-\frac{1}{2}\Delta \ln|\phi|^2=-\frac{1}{2}\Delta(\ln\kappa^2+ \ln|\phi|^2).
\ee

Set $u=\ln\kappa^2$ and $v=\ln|\phi|^2$ and note that $|\phi|$
behaves like $|x-p_s|$ for $x$ near $p_s$ ($s=1,\cdots,n$). We see
that $u$ and $v$ satisfy the equations
\begin{eqnarray}
 \label{o14}
  \Delta u &=&\lambda(2e^u-e^v-1),\\\label{o15}
  \Delta u+\Delta v &=& 2\lambda(e^v-1)+4\pi\sum^n_{s=1}
  \delta_{p_s}(x),
\end{eqnarray}

where we have included our consideration of the zero set $Z$ of
$\phi$ as given in (\ref{o6}).

\section {Solution on full plane}
\setcounter{equation}{0}\hspace*{1cm}

In this section, we prove the existence and uniqueness of the solution of the system of equations (\ref{o14}) and (\ref{o15}) over ${\mathbb{R}}^2$ satisfying the boundary condition
\be\label{21}
u\rightarrow 0, \hspace{0.5cm} v\rightarrow0 \hspace{0.5cm} \textmd{as} \quad |x|\rightarrow \infty .
\ee

To proceed further, we introduce the background function \cite{JT}
\be\label{22}
v_0(x)=-\sum^n_{s=1}\ln(1+\tau|x-p_s|^{-2}),\hspace{1cm}\tau > 0.
\ee

Then, we have
\be\label{23}
\Delta v_0=-h(x)+4\pi\sum^n_{s=1}\delta_{p_s}(x), \hspace{0.5cm}h(x)=4\sum^n_{s=1}\frac{\tau}{(\tau+|x-p_s|^2)^2} .
\ee
Using the substitution $v=v_0+w$, we have
\bea\label{24}
\Delta u &=& \lambda (2e^u-e^{v_0+w}-1), \\\label{25}
\Delta(u+w)&=& 2\lambda (e^{v_0+w}-1)+h(x).
\eea

Taking $f=u+w$, we change (\ref{24}) and (\ref{25}) into
\bea\label{26}
\Delta u &=& \lambda (2e^u-e^{v_0+f-u}-1), \\\label{27}
\Delta f &=& 2\lambda (e^{v_0+f-u}-1)+h(x).
\eea

It is clear that (\ref{26}) and (\ref{27}) are the Eulur--Lagrange  equations of the action functional
\bea\label{28}
  I(u,f) &=& \int_{{\mathbb{R}}^2}\biggl\{\frac{1}{2\lambda}|\nabla u|^2+\frac{1}{4\lambda}|\nabla f|^2+(2(e^u-1)-u)\nn\\
   &&+(e^{v_0+f-u}-e^{v_0})+(\frac{h}{2\lambda}-1)f\biggr\}dx.
\eea
It is clear that the  functional $I$ is a $C^1$-functional for $u,f\in W^{1,2}({\mathbb{R}}^2)$ and its Fr$\acute{e}$chet derivative  satisfies
\bea\label{29}
D I(u,f)(u,f) &=& \int_{{\mathbb{R}}^2}\biggl\{\frac{1}{\lambda}|\nabla u|^2+\frac{1}{2\lambda}|\nabla f|^2+ e^{v_0}(e^{f-u}-1)(f-u)\nn\\
  && +2(e^u-1)u+(e^{v_0}-1)(f-u)+\frac{h}{2\lambda}f\biggr\}dx.
\eea

Since
\be\label{210}
|\nabla u|^2+|\nabla f|^2=2|\nabla u|^2+|\nabla w|^2+2(\nabla u,\nabla w),
\ee
Hence
\be\label{211}
|\nabla u|^2+|\nabla f|^2\leq3|\nabla u|^2+2|\nabla w|^2\leq3(|\nabla u|^2+|\nabla w|^2).
\ee

On the other hand, we have
\bea\label{212}
 |\nabla u|^2+|\nabla f|^2 &\geq& 2|\nabla u|^2+|\nabla w|^2-2|(\nabla u,\nabla w)| \nn\\
   &\geq& (2-\frac{1}{\varepsilon})|\nabla u|^2+(1-\varepsilon)|\nabla w|^2,
\eea
for any $\varepsilon \in (\frac{1}{2},1)$.\\

Taking $\varepsilon =\frac{2}{3}$, we get
\be\label{213}
 |\nabla u|^2+|\nabla f|^2\geq\frac{1}{2}|\nabla u|^2+\frac{1}{3}|\nabla w|^2\geq\frac{1}{3}(|\nabla u|^2+|\nabla w|^2).
\ee

Similarly, we have
\be\label{214}
\frac{1}{3}(u^2+w^2)\leq u^2+f^2\leq3(u^2+w^2).
\ee
As a consequence of (\ref{29}), (\ref{213}) and (\ref{214}), we obtain
\bea\label{215}
 &&D I(u,f)(u,f)-\frac{1}{6\lambda} \int_{{\mathbb{R}}^2}\biggl\{|\nabla u|^2+|\nabla w|^2\biggr\}dx\nn\\
   &\geq&  \int_{{\mathbb{R}}^2}\biggl\{e^{v_0}(e^w-1)w+2(e^u-1)u+(e^{v_0}-1)w+\frac{h}{2\lambda}(u+w)\biggr\}dx \nn\\
   &=& \int_{{\mathbb{R}}^2}\biggl\{(e^{v_0}(e^w-1)+e^{v_0}-1+\frac{h}{2\lambda})w  +(2(e^u-1)+\frac{h}{2\lambda})u\biggr\}dx \nn \\
  &=& \int_{{\mathbb{R}}^2}\biggl\{w(e^{v_0+w}-1+\frac{h}{2\lambda}) +u(2(e^u-1)+\frac{h}{2\lambda})\biggr\}dx  \nn\\
  &\equiv& M_1(w)+M_2(u).
\eea
As in \cite{JT}, we decompose $w$ and $u$ into their positive and negative parts, $w=w_+-w_-$ and  $u=u_+-u_-$, where $q_+=\max\{q,0\}$ and $q_-=-\min\{q,0\}$ for $q\in {\mathbb{R}}$. Using the elementary inequality
\be\label{216}
e^t-1\geq t,\hspace{0.5cm}t\in {\mathbb{R}},
\ee
we have
\be\label{217}
e^{v_0+w}-1+\frac{h}{2\lambda}\geq v_0+w+\frac{h}{2\lambda} ,
\ee
which leads to
\bea\label{218}
 M_1(w_+)&\geq&\int_{{\mathbb{R}}^2}w^2_+dx+\int_{{\mathbb{R}}^2}w_+(v_0+\frac{h}{2\lambda})dx\nn\\
&\geq&\frac{1}{2}\int_{{\mathbb{R}}^2}w^2_+dx-\frac{1}{2}\int_{{\mathbb{R}}^2}(v_0+\frac{h}{2\lambda})^2dx.
\eea
On the other hand, using the inequality
\be\label{219}
1-e^{-t}\geq \frac{t}{1+t}, \hspace{0.5cm}t\geq 0,
\ee
we have
\bea\label{220}
   w_-(1-\frac{h}{2\lambda}-e^{v_0-w_-})
   &=& w_-(1-\frac{h}{2\lambda}+e^{v_0}(1-e^{-w_-})-e^{v_0}) \nn\\
   &\geq& w_-(1-\frac{h}{2\lambda}+e^{v_0}\frac{w_-}{1+w_-}-e^{v_0}) \nn\\
   &=&  \frac{w_-^2}{1+w_-}(1-\frac{h}{2\lambda})+ \frac{w_-}{1+w_-}(1-e^{v_0}-\frac{h}{2\lambda}).
\eea
In view of (\ref{23}), we see that we may choose $\tau>0$ large  enough so that
\be\label{221}
\frac{h(x)}{\lambda}<1, \hspace{0.5cm}x\in {\mathbb{R}}^2.
\ee
Since $1-e^{v_0}$ and $h$ both lie in $L^2({\mathbb{R}}^2)$, we have
\be\label{222}
  \int_{{\mathbb{R}}^2}\frac{w_-}{1+w_-}\biggl|1-e^{v_0}-\frac{h}{2\lambda}\biggr|dx \leq \varepsilon \int_{{\mathbb{R}}^2}\frac{w_-^2}{1+w_-}dx+C(\varepsilon),
\ee
where $\varepsilon> 0$ may be  chosen to be arbitrarily small. Combining (\ref{220})-(\ref{222}), we obtain
\be\label{223}
  M_1(-w_-) \geq \frac{1 }{4}\int_{{\mathbb{R}}^2}\frac{w_-^2}{1+w_-}dx-C_1(\varepsilon),
\ee
provided that $\varepsilon< \frac{1 }{4}$. From (\ref{218}) and (\ref{223}), we get
\be\label{224}
 M_1(w) \geq  \frac{1 }{4}\int_{{\mathbb{R}}^2}\frac{w^2}{1+|w|}dx-C,
\ee
where and  in the sequel we use $C$ to denote an irrelevant positive constant.
Similar estimates may be made for $M_2(u)$. Thus, (\ref{215}) gives us
\bea\label{225}
 && D I(u,f)(u,f)-\frac{1}{6\lambda} \int_{{\mathbb{R}}^2}\biggl\{|\nabla u|^2+|\nabla w|^2\biggr\}dx\nn \\
 &\geq& \frac{1 }{4}\int_{{\mathbb{R}}^2}\biggl(\frac{u^2}{1+|u|}+\frac{w^2}{1+|w|}\biggr)dx-C.
\eea We now recall the well-known Gagliardo--nirenberg inequality
\be\label{226}
 \int_{{\mathbb{R}}^2}u^4dx\leq 2\int_{{\mathbb{R}}^2}u^2dx \int_{{\mathbb{R}}^2}|\nabla u|^2dx,\quad u\in W^{1,2}({\mathbb{R}}^2).
\ee
Consequently, we have
\bea\label{227}
\biggl(\int_{{\mathbb{R}}^2}u^2dx\biggr)^2 &=& \biggl(\int_{{\mathbb{R}}^2}\frac{|u|}{1+|u|}(1+|u|)|u|dx\biggr)^2\nn \\
   &\leq& \int_{{\mathbb{R}}^2}\frac{u^2}{(1+|u|)^2}dx\int_{{\mathbb{R}}^2}(1+|u|)^2|u|^2dx\nn \\
   &\leq& 2 \int_{{\mathbb{R}}^2}\frac{u^2}{(1+|u|)^2}dx\int_{{\mathbb{R}}^2}(u^2+u^4)dx \nn\\
   &\leq& 4 \int_{{\mathbb{R}}^2}\frac{u^2}{(1+|u|)^2}dx\int_{{\mathbb{R}}^2}u^2dx\biggl(1+\int_{{\mathbb{R}}^2}|\nabla u|^2dx\biggr) \nn \\
   &\leq& \frac{1}{2}\biggl(\int_{{\mathbb{R}}^2}u^2dx \biggr)^2+C\biggl(1+[\int_{{\mathbb{R}}^2}\frac{u^2}{(1+|u|)^2}dx]^4\nn\\
   &&+[\int_{{\mathbb{R}}^2}|\nabla u|^2dx]^4\biggr).
\eea
As a result of (\ref{227}), we have
\be\label{228}
  \biggl(\int_{{\mathbb{R}}^2}u^2dx\biggr)^{\frac{1}{2}} \leq C\biggl(1+\int_{{\mathbb{R}}^2}|\nabla u|^2dx+\int_{R^2}\frac{u^2}{(1+|u|)^2}dx\biggr) .
\ee
Applying (\ref{228}) in (\ref{225}), we arrive at
\be\label{229}
 D I(u,f)(u,f)\geq C_1(\|u\|_{W^{1,2}({\mathbb{R}}^2)}+\|w\|_{W^{1,2}({\mathbb{R}}^2)}) -C_2.
\ee
Thus, using (\ref{211}), (\ref{213}) and (\ref{214}) in (\ref{229}), we conclude with the coercive lower bound
\be\label{230}
 D I(u,f)(u,f)\geq C_1(\|u\|_{W^{1,2}({\mathbb{R}}^2)}+\|f\|_{W^{1,2}({\mathbb{R}}^2)}) -C_2.
\ee

With (\ref{230}), we can now show that the existence of a critical point of the action functional (\ref{28}) follows by using a standard argument as in \cite{YA}.

In fact, from (\ref{230}), we can choose $R>0$ large enough such that
\be\label{231}
\inf\{D I(u,f)|u,f\in W^{1,2}({\mathbb{R}}^2),\|u\|_{W^{1,2}({\mathbb{R}}^2)}+\|f\|_{W^{1,2}({\mathbb{R}}^2)}=R\}\geq 1
\ee
(say). Now consider the minimization problem
\be\label{232}
  \eta =\min\{I(u,f)|\|u\|_{W^{1,2}({\mathbb{R}}^2)}+\|f\|_{W^{1,2}({\mathbb{R}}^2)}\leq R\}.
\ee

Let $\{(u_k,f_k)\}$ be a  minimization sequence of (\ref{232}). Without  loss of generality, we may assume that $\{(u_k,f_k)\}$ weakly converges to an element $(u,f)$  in $W^{1,2}({\mathbb{R}}^2)$. The weakly lower semi-continuity of $I$ implies that $(u,f)$ solves (\ref{232}). To show that $(u,f)$  is a critical pint of $I$, it suffices to see that it is an interior point. That is,
\be\label{233}
  \|u\|_{W^{1,2}({\mathbb{R}}^2)}+\|f\|_{W^{1,2}({\mathbb{R}}^2)}< R.
\ee
Suppose otherwise that $  \|u\|_{W^{1,2}({\mathbb{R}}^2)}+\|f\|_{W^{1,2}({\mathbb{R}}^2)}= R$. Then for $t\in(0,1)$ the point $(1-t)(u,f)$
is interior which gives us
\be\label{234}
  I((1-t)u,(1-t)f) \geq \eta=I(u,f).
\ee
On the other hand, we have
\bea\label{235}
  \lim_{t\rightarrow0}\frac{I((1-t)(u,f))-I(u,f)}{t} &=& \frac{d}{dt}I((1-t)(u,f))|_{t=0}\nn \\
   &=& - D I(u,f)(u,f)\leq -1.
\eea
Consequently, if $t>0$ is sufficiently small, (\ref{234}) leads to
\be\label{236}
  I((1-t)(u,f))<I(u,f)=\eta,
\ee
which contradicts (\ref{234}). Therefore, the existence of a critical point of $I$ follows.

Note that the part in the integrand  of $I$ which does not involve the derivatives of $u$ and $f$ may be rewritten
as
\be\label{237}
  Q(u,f) =2(e^u-1)-u +e^{v_0+f-u}-e^{v_0}+(\frac{h}{2\lambda}-1)f ,
\ee
whose Hessian is easily checked to be positive definite. Thus, the functional $I$ is strictly convex. As a consequence, $I$ can have at most  one critical point $(u,f)$ in the space $W^{1,2}({\mathbb{R}}^2)$.

To proceed further, we now show that the following claim holds.

\textbf{Claim}: If $g\in W^{1,2}({\mathbb{R}}^2)$, then $e^g-1\in L^2({\mathbb{R}}^2)$.

We first recall the Sobolev embedding inequality in two dimensions \cite{GT}:
\be\label{238}
  \|g\|_{L^k({\mathbb{R}}^2)} \leq \biggl(\pi (\frac{k-2}{2})\biggr)^{\frac{k-2}{2k}}\|g\|_{W^{1,2}({\mathbb{R}}^2)},\quad k\geq 2.
\ee
On the other hand, the MacLaurin series leads to
\be\label{239}
  (e^g-1)^2 =g^2+\sum^\infty_{k=3}\frac{2^k-2}{k!}g^k.
\ee
Combining the above with (\ref{238}), we have, formally,
\be\label{240}
  \|e^g-1\|^2_{L^2({\mathbb{R}}^2)} \leq  \|g\|^2_{L^k({\mathbb{R}}^2)}+\sum^\infty_{k=3}\frac{2^k-2}{k!}\biggl(\pi \frac{k-2}{2}\biggr)^{\frac{k-2}{2}}\|g\|^k_{W^{1,2}({\mathbb{R}}^2)}.
\ee
Setting $$\alpha_k=\frac{2^k-2}{k!}\biggl(\pi \frac{k-2}{2}\biggr)^{\frac{k-2}{2}}\|g\|^k_{W^{1,2}({\mathbb{R}}^2)},$$
and applying the Stirling formula,
\be\label{241}
  k!\sim\sqrt{2\pi}k^{k+\frac{1}{2}}e^{-k}\quad (k\rightarrow \infty),
\ee
we have
\bea\label{242}
 \sqrt[k]{\alpha _k}&\sim&\frac{\sqrt[k]{2^k-2}}{ke^{-1}(2k\pi)^{\frac{1}{2k}}}\biggl(\pi \frac{k-2}{2}\biggr)^{\frac{k-2}{2k}}\|g\|_{W^{1,2}({\mathbb{R}}^2)}\nn\\
 &\sim&2e\sqrt{\pi}\|g\|_{W^{1,2}({\mathbb{R}}^2)}\biggl(\frac{k-2}{2k^2}\biggr)^{\frac{1}{2}}\rightarrow0<1\quad (k\rightarrow \infty).
\eea
Thus we have shown that (\ref{240}) is a convergent series, which verifies our claim.

We now continue our work. Noting $v_0,h \in L^2({\mathbb{R}}^2)$ and  using the claim, we see that the right-hand side of (\ref{26}) and (\ref{27}) belong to $L^2({\mathbb{R}}^2)$. We may now apply the standard elliptic theory to (\ref{24}) and (\ref{25}) to infer that $u,w \in W^{2,2}({\mathbb{R}}^2)$. In particular, $u, w$ and $|\nabla u|, |\nabla w|$ approach zero as $|x|\rightarrow\infty$, which renders the validity of the boundary condition (\ref{21}).

Finally, we derive the decay rates for $u, v$ and $|\nabla u|, |\nabla v|$. Consider (\ref{o14}) and (\ref{o15}) outside the disk $D_{R}=\biggr\{x\in {\mathbb{R}}^2\biggr||x|< R\biggl\}$, where
$$R>\max\biggr\{|p_s|\biggr|s=1,2,\cdots,n\biggl\}.$$
We rewrite (\ref{o14}) and (\ref{o15}) in ${\mathbb{R}}^2\setminus
D_{R}$ as \bea\label{243} \Delta
u&=&\lambda(2e^u-e^v-1),\\\label{244} \Delta
v&=&\lambda(-2e^u+3e^v-1). \eea

By computation, we have
\bea\label{245}
\Delta(u^2+v^2)&=& 2(|\nabla u|^2+ |\nabla v|^2)+4\lambda u(e^u-1)+6\lambda v(e^v-1)\nn\\
&&-2\lambda u(e^v-1)-4\lambda v(e^u-1),\quad
x\in{\mathbb{R}}^2\setminus D_{R}. \eea Noting $u,v\rightarrow 0$ as
$|x|\rightarrow \infty$, for any $\varepsilon: 0<\varepsilon<1$, we
can find a suitably large $R_\varepsilon>R$ so that \bea\label{246}
\Delta(u^2+v^2)&\geq&4\lambda u^2+6\lambda v^2-3(2+\varepsilon)\lambda|uv|\nn\\
&\geq&(1-\varepsilon)\lambda(u^2+v^2),\quad
x\in{\mathbb{R}}^2\setminus D_{R_\varepsilon}. \eea Thus, using a
comparison function argument and the property $u^2+v^2=0$ at
infinity, we can obtain a constant $C(\varepsilon)>0$ to make
\be\label{247} u^2(x)+v^2(x)\leq
C(\varepsilon)e^{-\sqrt{(1-\varepsilon)\lambda}|x|} \ee valid.

Let $\partial$ denote any of the two partial derivatives, $\partial_1$ and $\partial_2$. Then (\ref{o14}) and (\ref{o15}) yields
\bea\label{248}
\Delta (\partial u)&=& \lambda (2(\partial u)e^u-(\partial v)e^v),\\\label{249}
\Delta (\partial v)&=& 3\lambda(\partial v)e^v-2\lambda(\partial u)e^u.
\eea

By computation and then using the Canchy inequality, we get
\bea\label{250}
\Delta(|\nabla u|^2+|\nabla v|^2)&=&2[|\nabla(\partial_ 1u)|^2+|\nabla (\partial_2u)|^2+|\nabla(\partial_ 1v)|^2+|\nabla (\partial_2v)|^2]\nn\\
&&+4\lambda|\nabla u|^2e^u+6\lambda|\nabla v|^2e^v-2\lambda\partial_1u\partial_1ve^v-2\lambda\partial_2u\partial_2ve^v\nn\\
&&-4\lambda\partial_1u\partial_1ve^u-4\lambda\partial_2u\partial_2ve^u\nn\\
&\geq& 4\lambda |\nabla u|^2e^u+6\lambda|\nabla v|^2e^v-\lambda|\nabla u|^2(1+2e^{2u})\nn\\
&&-\lambda|\nabla v|^2(2+e^{2v}),\quad x\in{\mathbb{R}}^2\setminus
D_{R}. \eea Therefore, as before, we conclude that for any
$\varepsilon: 0<\varepsilon<1$, there is a
$\widetilde{R}_\varepsilon> R$, so that \be\label{251}
\Delta(|\nabla u|^2+|\nabla v|^2)\geq2(1-\varepsilon)(|\nabla
u|^2+|\nabla v|^2),\quad x\in{\mathbb{R}}^2\setminus
D_{\widetilde{R}_\varepsilon}. \ee Noting the property $|\nabla
u|^2+|\nabla v|^2=0$ at infinity, applying the comparison principle,
we arrive at \be\label{252} |\nabla u|^2+|\nabla v|^2\leq
C(\varepsilon)e^{-\sqrt{2(1-\varepsilon)\lambda}|x|}, \quad |x|>R.
\ee Inserting this information into (\ref{24}) and (\ref{25}), we
see that the associated functions $u, w$ and the right-hand sides of
(\ref{24}) and (\ref{25}) all lie in $L^2({\mathbb{R}}^2)$.
Consequently, the pair of functions $u$ and $f$ yields a
$W^{2,2}({\mathbb{R}}^2)$-solution of (\ref{26}) and (\ref{27}),
which must be the unique critical point of the functional $I$
produced earlier.

We may summarize our results as follows.

\begin{theorem}\label{theorem031}  For any distribution of the points $p_1,p_2, \cdots p_n \in {\mathbb{R}}^2$, the system of nonlinear elliptic equations (\ref{o14}) and (\ref{o15}) subject to the boundary condition (\ref{21})  has a unique solution. Furthermore, the solution satisfy (\ref{247}) and (\ref{252}) decay estimates at infinity.
\end{theorem}

\section {Solution via variational approach on doubly periodic domain}
\setcounter{equation}{0} \hspace*{1cm}

In this section, we consider solutions of (\ref{o14}) and (\ref{o15}) defined over a
doubly periodic domain $\Omega$. In order to get rid of the singular
source terms, we introduce a background function $v_0$ satisfying
\be
  \Delta v_0 = -\frac{4\pi n}{|\Omega|}+4\pi\sum^n_{s=1} \delta_{p_s}(x).
\ee
Using the new variable $w$ so that $v=v_0+w$, we can modify
(\ref{o14}) and (\ref{o15}) into \bea\label{32}
  \Delta u &=&\lambda(2e^u-e^{v_0+w}-1) ,\\\label{33}
  \Delta u+\Delta w &=& 2\lambda(e^{v_0+w}-1)+\frac{4\pi n}{|\Omega|}.
\eea
Note that, since the singularity of $v_0$ at $p_s$ is of the
type $\ln|x-p_s|^2$, the weight function $e^{v_0}$ is everywhere
smooth.

To proceed further, we take $u+w=f$. Then the governing system of  equations become
\bea\label{34}
  \Delta u &=& \lambda(2e^u-e^{v_0+f-u}-1),\\\label{35}
  \Delta f &=& 2\lambda(e^{v_0+f-u}-1) +\frac{4\pi n}{|\Omega|}.
\eea

Integrating (\ref{35}) and (\ref{34}), we have
\bea\label{36}
  \int_\Omega e^{v_0+f-u} dx&=& |\Omega| -\frac{2\pi
n}{\lambda}\equiv
  C_1>0,\\
  \label{37}
  \int_\Omega e^u dx &=& \frac{1}{2}\int_\Omega e^{v_0+f-u} dx+\frac{1}{2}|\Omega|=\frac{1}{2}(C_1+|\Omega|)\equiv C_2>0,
\eea Of course, the conditions (\ref{36}) and (\ref{37}) imply that
the existence of an n-vortex solution requires that $C_{1}>0$ and
$C_{2}>0$, which is simply
\be\label{38}
    |\Omega|-\frac{2\pi
        n}{\lambda}\equiv C_{1}>0,
\ee since $C_{1}>0$ contains $C_{2}>0$.

We can prove that (\ref{38}) is in fact sufficient for existence as
well.

\begin{theorem}\label{theorem31}   The system of the non-Abelian vortex equations
(\ref{34}) and (\ref{35}) has a solution if and only if (\ref{38})
holds or \be\label{39}
    2\pi n<\lambda|\Omega|.
\ee Furthermore, if a solution exists, it must be unique, which can
be constructed through solving a multiply constrained minimization
problem.
\end{theorem}

We use $W^{1,2}(\Omega)$ to denote the usual Sobolev space of doubly periodic functions over the cell domain $\Omega$. We will prove Theorem \ref{theorem31} in terms of three lemmas as follows:

\begin{lemma}\label{lemm31} Consider the constrained minimization problem
\be\label{310}
  \min \{I(u,f)|u,f\in W^{1,2}(\Omega),J_k(u,f)=C_k,k=1,2\},
\ee where
\bea
  I(u,f) &=& \int _\Omega \biggl\{\frac{1}{2}|\nabla u |^2+\frac{1}{4}|\nabla f |^2-\lambda u-\lambda f+\frac{2\pi n}{|\Omega |}f\biggr\}\dd x,\\
 J_1(u,f) &=& \int_\Omega e^{v_0} e^{f-u}\dd x=C_1 ,\\
 J_2(u,f)&=& \int_\Omega e^ u \dd x=C_2.
\eea Then a solution of (\ref{310}) is a solution of the system of  equations (\ref{34}) and (\ref{35}).
\end{lemma}
\begin{proof}
It is clear that the Fr$\acute{e}$chet derivatives $d J_1, d J_2$ of
the constraint functionals are linearly independent.

Let $(u,f)$ be a solution of (\ref{310}). Then by standard elliptic
regularity theory $(u, f)$ must be smooth and there exist Lagrange
multipliers $\lambda _1, \lambda_2\in {\mathbb{R}}$ so that
\bea\label{314}
  \Delta u&=&-\lambda -\lambda _1 e^{v_0} e^{f-u} +\lambda _2 e^u,\\\label{315}
  \Delta f&=&-2\lambda +2\lambda _1e^{v_0} e^{f-u}+\frac{4\pi n}{|\Omega |}.
\end{eqnarray}
Integrating the equation (\ref{315}) and using $J_1 (u,f)=C_1$, we
obtain $\lambda _1=\lambda$ which means that $(u,f)$ verifies the
equation (\ref{35}). To recover the equation (\ref{34}), we use $J_2(u,f)=C_2$. By virtue of $\lambda _1=\lambda $ and integrating the equation (\ref{314}), we have $\lambda _2=2\lambda $.

In particular, $(u,f)$  is the solutions of the equations
(\ref{34}), (\ref{35}). The lemma is proven.
\end{proof}

The admissible set of the variation problem (\ref{310}) will be
denoted by
\be\label{316}
  {\mathcal{C}} =\{u,f\in  W^{1,2}(\Omega)| J_k(u,f)=C_k,k=1,2\} .
\ee
When (\ref{36}) and (\ref{37}) are satisfied, $C_1, C_2>0$. Thus $\mathcal{C}\neq \emptyset$.

\begin{lemma}\label{lemm32}
If the condition (\ref{39}) holds, then (\ref{310}) has a solution.
In other words, the system (\ref{o14})-(\ref{o15}) has a solution if
and only if (\ref{39}) is fulfilled.
\end{lemma}
\begin{proof}
By virtue of lemma \ref{lemm31}, it is sufficient to show the existence of a minimizer of the constrained optimization problem (\ref{310}).

We first proved that under the condition (\ref{38}) or (\ref{39}),
the objective functional $I$ is bounded from below on ${\mathcal{C}}$.
For this purpose, we rewrite each $\eta\in W^{1,2}(\Omega)$ as follows
\be
 \eta=\underline{\eta}+{\eta}',\nn
\ee
where $\underline{\eta}$ denotes the integral mean of $\eta$,
$\underline{\eta}=\frac{1}{|\Omega|}\int _\Omega \eta dx$ and $\int_\Omega \eta' dx=0$. Hence, $I$ may be put for $(u, f)\in{\mathcal{C}}$ in the form
\be\label{317}
  I(u,f)=\int _\Omega \{\frac{1}{2} |\nabla u'|^2 +\frac{1}{4} |\nabla f'|^2\}dx-\lambda \underline{f}|\Omega|+2\pi n \underline{f} -\lambda \underline{u}|\Omega|.
\ee
Setting
\be\label{318}
  \Lambda(\underline{u}, \underline{f}) =-(\lambda|\Omega|-2\pi n)\underline{f}-\lambda |\Omega |\underline{u},
\ee we can derive from (\ref{36}) and (\ref{37}) the expressions
 \bea\label{319}
  {\underline{u}}&=&\ln {C_2}-\ln\biggl(\int_{\Omega} e^{u'}\biggr),\\
\label{320}
 \underline{f}&=&\ln (C_1C_2)-\ln\biggl(\int_{\Omega} e^{u'}\biggr)-\ln\biggl(\int_{\Omega} e^{v_0+f'-u'}\biggr).
 \eea
Inserting (\ref{319}) and (\ref{320}) into (\ref{318}), we have
$$
\Lambda(\underline{u},\underline{f})=(2\lambda |\Omega|-2\pi
      n)\ln\biggl(\int_{\Omega} e^{u'}\biggr)+(\lambda |\Omega|-2\pi n)\ln\biggl(\int_{\Omega} e^{v_0+f'-u'}\biggr)+C_3,
$$
where $C_3=(2\pi n-\lambda |\Omega|)\ln C_1+(2\pi n-2\lambda |\Omega|)\ln C_2$.

Using  Jensen's inequality, we get
\begin{eqnarray*}
  \ln \biggl[\int_{\Omega} \exp(v_0+f'-u')\biggr]
  &\geq & \ln\biggl[|\Omega |\exp\biggl(\frac{1}{|\Omega |}\int_{\Omega} (v_0+f'-u')\biggr)\biggr]\\
  &=&\ln \biggl[|\Omega|\exp\biggl(\frac{1}{|\Omega |}\int_{\Omega} v_0 \biggr)\biggr],\\
  \ln \biggl[\int_{\Omega}{e^{u'}}\biggr]&\geq &\ln \biggl [|\Omega |\exp\biggl(\frac{1}{|\Omega|}\int_{\Omega} u'\biggr)\biggr]=\ln {|\Omega |}.
\end{eqnarray*}
Noting  (\ref{39}), we have
\bea\label{321}
  \Lambda(\underline{u},\underline{f})&\geq& (2\lambda |\Omega|-2\pi
      n)\ln {|\Omega |}+(\lambda |\Omega|-2\pi n)\ln \biggl[|\Omega|\exp\biggl(\frac{1}{|\Omega |}\int_{\Omega} v_0  \biggr )\biggr]\nn\\
      &&+C_3.
 \eea
Inserting (\ref{321}) into (\ref{317}), we arrive at the coercive lower estimate
\be\label{322}
 I(u,f)\geq \int_{\Omega}\biggl\{\frac{1}{2} |\nabla u'|^2+\frac{1}{4}|\nabla f'|^2\biggr\}dx-C_4,
\ee
where $C_4>0$ is an irrelevant constant. From (\ref{322}), we know that the existence of solution of (\ref{310}) follows.

In fact, let $\{(u_j, f_j)\}\subset \mathcal{C}$ be a minimizing
sequence of the variational problem (\ref{310}) and set
\be\label{323} \underline{f}_j=\frac{1}{|\Omega|}\int_{\Omega}f_j
dx,\quad\underline{u}_j=\frac{1}{|\Omega|}\int_{\Omega}u_j dx. \ee
Then, with $u'_j=u_j-\underline{u}_j$ and
$f'_j=f_j-\underline{f}_j$, we have $\underline{u}'_j=0$ and
$\underline{f}'_j=0$. In view of (\ref{322}), we see that
$\{(u'_j,f'_j)\}$ is bounded in $W^{1,2}(\Omega)$. Without loss of
generality, we may assume that $\{(u'_j,f'_j)\}$ converges weakly in
$W^{1,2}(\Omega)$ to an element $(u',f')$ (say). The compact
embedding \be\label{324} W^{1,2}(\Omega)\hookrightarrow
L^p(\Omega),\quad p\geq 1, \ee then implies
$(u'_j,f'_j)\rightarrow(u',f')$ in $L^p(\Omega)~(p\geq 1)$ as
$j\rightarrow\infty$. In particular, $\underline{u}'=0$ and
$\underline{f}'=0$.

Recall the Trudinger-Moser inequality \cite{A}
\be\label{325}
\int_\Omega e^F dx\leq C(\varepsilon) \exp\biggl(\biggl[\frac{1}{16\pi}+\varepsilon\biggr]\int_\Omega |\nabla F|^2 dx\biggr),\quad F\in W^{1,2}(\Omega),\quad \underline{F}=0.
\ee
where $C(\varepsilon)>0$ is a constant. In view of (\ref{324}) and (\ref{325}), we see that the functionals defined by the right-hand side of (\ref{319}) and (\ref{320}) are continuous in $u', f'$ with respect to the weak topology of $W^{1,2}(\Omega)$. Therefore, $\underline{u}_j\rightarrow\underline{u},\underline{f}_j\rightarrow\underline{f}$ as $j\rightarrow \infty$, as given in (\ref{319}) and (\ref{320}). In other words, $(u,f)=(\underline{u}+u',\underline{f}+f')$ satisfies the constraints (\ref{36}) and (\ref{37}), and solves the constrained minimization problem (\ref{310}). Thus Lemma \ref{32} is proven.
\end{proof}

Now we state the uniqueness of the solution to the equations (\ref{34})
and (\ref{35}) as follows.

\begin{lemma}\label{lemm33} If system (\ref{34})-(\ref{35}) has a solution, then the solution
must be unique.
\end{lemma}
\begin{proof}
Consider the following functional,
\begin{eqnarray*}
  J(u,f)&=&\frac{1}{2}||\nabla u||^2_2+\frac{1}{4}||\nabla f||^2_2+(-\lambda \underline{u}-\lambda\underline{f}+\frac{2\pi n}{|\Omega|}\underline{f})|\Omega|\\
  &&+\int_ \Omega\{\lambda e^{v_0+f-u}+2\lambda e^u\}dx.
\end{eqnarray*}
It is straightforward to check by calculating the Hessian that $J$
is strictly convex in $W^{1,2}(\Omega)$. Thus $J$ has at most one critical point.
However, any solution of (\ref{34})-(\ref{35}) must be a critical
point of $J$. This proves the lemma.
\end{proof}

\section {Solution via fixed point method on doubly periodic domain}
\setcounter{equation}{0} \hspace*{1cm}

In this section, we shall solve the problem (\ref{o14}) and (\ref{o15})
by a fixed-point method via the Leray--Schauder theorem. This approach is of independent interest because the a priori estimates obtained in the process provide additional information on the governing equations.

We rewrite (\ref{o14}) and (\ref{o15}) as
\bea\label{41}
  \Delta u &=& \lambda(2e^u-e^v-1) ,\\\label{42}
   \Delta v &=&\lambda(-2e^u+3e^v-1)+4\pi\sum^n_{s=1}\delta_{p_s}(x).
\eea
From Section 4, we know that (\ref{36}) and (\ref{37}) are a necessary condition for the solvability of (\ref{41}) and (\ref{42}) over a doubly periodic domain $\Omega$.

We now proceed to prove (\ref{36}) and (\ref{37}) are also sufficient for the existence of a solutions to the equations (\ref{41}) and (\ref{42}). Using a fixed point argument over the Sobolev space $W^{1,2}(\Omega)$. Consider the proper subspace of $W^{1,2}(\Omega)$ defined by
\be\label{43}
X=\biggl\{(u',v')\in W^{1,2}(\Omega)\biggl|\int_\Omega u'dx=0, \int_\Omega v'dx=0\biggr\},
\ee
where $u'=u-\underline{u}, v'=v-\underline{v}$, and
\bea\label{45}
  \underline{v}&=&\ln C_1-\ln\biggl( \int_\Omega e^{v'}dx\biggr),\\
 \label{46}
   \underline{u}&=&\ln C_2-\ln\biggl( \int_\Omega e^{u'}dx\biggr),
\eea
after resolving the constraints (\ref{36}) and (\ref{37}). By the Poincar$\acute{e}$ inequality \cite{M}, we may define the norm of $X$ as follow
\be\label{44}
\|(u',v')\|_X=\|\nabla u'\|_{L^2(\Omega)}+\|\nabla v'\|_{L^2(\Omega)}.
\ee
For each given $(u',v')\in X$, consider the equations
\begin{eqnarray} \label{47}
 \Delta U' &=& \lambda\biggl(\frac{2C_2e^{u'}}{\int_\Omega e^{u'}dx}-\frac{C_1e^{v'}}{\int_\Omega e^{v'}dx }-1\biggr), \\ \label{48}
  \Delta V' &=& \lambda\biggl(\frac{-2C_2e^{u'}}{\int_\Omega e^{u'}dx}+\frac{3C_1e^{v'}}{\int_\Omega e^{v'}dx }-1\biggr)+4\pi\sum^n_{s=1}\delta_{p_s}(x).
\end{eqnarray}
By (\ref{36}) and (\ref{37}), we see that the right-hand side of
(\ref{47}) and (\ref{48}) have zero average value on $\Omega$.
Therefore the equations (\ref{47}) and (\ref{48}) has a unique
solution $(U',V')\in X$. This correspondence,
$(u',v')\longmapsto(U',V')$, gives us a well-defined operator $T$ that maps $X$ into itself, $(U',V')=T(u',v')$.

\begin{theorem}\label{theorem041} The system of equation (\ref{47}) and (\ref{48}) has a solution if and only if the conditions (\ref{36}) and (\ref{37}) are valid.
\end{theorem}
We will prove Theorem \ref{theorem041} in terms of two lemmas as follows:
\begin{lemma}\label{lemm41}
 The operator $T: X \longmapsto X$ is completely continuous.
\end{lemma}

\begin{proof}
 Let $(u'_n,v'_n)\rightarrow(u'_0,v'_0)$ weakly in $X$ as $n\rightarrow \infty$. Then $(u'_n,v'_n)\rightarrow
 (u'_0,v'_0)$ strongly in $L^p(\Omega)~(p\geq 1)$. Set $(U'_n,V'_n)=T(u'_n,v'_n)$ and
$(U'_0,V'_0)=T(u'_0,v'_0)$. Then \bea \label{49}
  \Delta (U'_n-U'_0)&=& \lambda \biggl(\frac{2C_2e^{u'_n}}{\int_\Omega e^{u'_n}dx}-\frac{C_1e^{v'_n}}{\int_\Omega e^{v'_n}dx}-\frac{2C_2e^{u'_0}}{\int_\Omega e^{u'_0}dx}+\frac{C_1e^{v'_0}}{\int_\Omega e^{v'_0}dx}\biggr), \\
  \label{410}\Delta (V'_n-V'_0) &=& \lambda\biggl( \frac{-2C_2e^{u'_n}}{\int_\Omega e^{u'_n}dx}+\frac{3C_1e^{v'_n}}{\int_\Omega e^{v'_n}dx}+\frac{2C_2e^{u'_0}}{\int_\Omega e^{u'_0}dx}-\frac{3C_1e^{v'_0}}{\int_\Omega e^{v'_0}dx}\biggr).
\eea
Multiplying  (\ref{49}) and (\ref{410}) by $U'_n-U'_0$  and $V'_n-V'_0$, and integrating by parts, respectively, we obtain

\bea\label{411}
  \int_\Omega| \nabla (U'_n-U'_0)|^2dx &=&\int_\Omega\lambda \biggl\{\frac{2C_2e^{u'_n}}{\int_\Omega e^{u'_n}dx}-\frac{2C_2e^{u'_0}}{\int_\Omega e^{u'_0}dx}\nn\\
  &&+ \frac{C_1e^{v'_n }}{\int_\Omega e^{v'_n}dx}\ - \frac{C_1e^{v'_0}}{\int_\Omega e^{v'_0}dx}\biggr\}(U'_n-U'_0) dx, \\\label{412}
  \int_\Omega|\nabla (V'_n-V'_0)|^2 dx&=&\int_\Omega\lambda\biggl\{\frac{-2C_2e^{u'_n}}{\int_\Omega e^{u'_n}dx}+\frac{2C_2e^{u'_0}}{\int_\Omega e^{u'_0}dx}\nn\\
  &&+ \frac{3C_1e^{v'_n}}{\int_\Omega e^{v'_n}dx}\ - \frac{3C_1e^{v'_0}}{\int_\Omega e^{v'_0}dx}\biggr\}(V'_n-V'_0)dx.
\eea
Note that the boundedness of $\{(u'_n,v'_n)\}$ in $X$ and the Trudinger-Moser inequality \cite{A} imply that
\bea\label{413}
\sup_n\int_\Omega e^{u'_n}dx\leq C<\infty,\\
\label{414}\sup_n\int_\Omega e^{v'_n}dx\leq C<\infty.
\eea
 Therefore, from (\ref{411}), we obtain
\bea\label{415}
 \int_\Omega|\nabla(U'_n-U'_0)|^2dx
&\leq&\lambda\biggl\{\frac{4C_2}{\int_\Omega e^{u'_n}dx}\int_\Omega e^{\tilde{u}'_n}|u'_n-u'_0||U'_n-U'_0|dx\nn \\
&&+\frac{2C_1}{\int_\Omega e^{v'_n}dx}\int_\Omega e^{\tilde{v}'_n}|v'_n-v'_0||U'_n-U'_0|dx\biggr\}\nn\\
&\leq&\lambda\biggl\{\frac{4C_2}{|\Omega|}\int_\Omega e^{\tilde{u}'_n}|u'_n-u'_0||U'_n-U'_0|dx\nn\\
&&+\frac{2C_1}{|\Omega|}\int_\Omega e^{\tilde{v}'_n}|v'_n-v'_0||U'_n-U'_0|dx\biggr\},
\eea
where $\tilde{u}'_n$ and $\tilde{v}'_n$ lie between $u'_n,v'_n$ and $u'_0,v'_0$, respectively. In (\ref{415}), we have used the  inequalities
$$
\int_\Omega e^{u'_n}dx\geq|\Omega|\exp\biggl(\frac{1}{|\Omega|}\int_\Omega u'_n dx\biggr)=|\Omega|,
$$
and
$$
\int_\Omega e^{v'_n}dx\geq|\Omega|\exp\biggl(\frac{1}{|\Omega|}\int_\Omega v'_n dx\biggr)=|\Omega|.
$$
Applying the Canchy inequality and H$\ddot{o}$lder inequality, and (\ref{413}), we have
\bea\label{416}
\int_\Omega e^{\tilde{u}'_n}|u'_n-u'_0||U'_n-U'_0|dx
&\leq&\frac{1}{2\varepsilon}\int_\Omega e^{2\tilde{u}'_n}|u'_n-u'_0|^2dx+\frac{\varepsilon}{2}\int_\Omega|U'_n-U'_0|^2dx\nn\\
&\leq&\frac{1}{2\varepsilon}\biggl(\int_\Omega e^{4\tilde{u}'_n}dx\biggr)^{\frac{1}{2}}\biggl(\int_\Omega|u'_n-u'_0|^4x\biggr)^{\frac{1}{2}}\nn\\
&&+\frac{C_3\varepsilon}{2}\|\nabla(U'_n-U'_0)\|^2_{L^2(\Omega)}\nn\\
&\leq&C_\varepsilon\|u'_n-u'_0\|^2_{L^4(\Omega)}+\frac{C_3\varepsilon}{2}\|\nabla(U'_n-U'_0)\|^2_{L^2(\Omega)}.\nn\\
\eea
Similarly,
\be\label{417}
\int_\Omega e^{\tilde{v}'_n}|v'_n-v'_0||U'_n-U'_0|dx
\leq C_\varepsilon\|v'_n-v'_0\|^2_{L^4(\Omega)}+\frac{C_4\varepsilon}{2}\|\nabla(U'_n-U'_0)\|^2_{L^2(\Omega)}.
\ee
Inserting (\ref{416}) and (\ref{417}) into (\ref{415}), and letting $\varepsilon> 0$ be small enough, we have
\be\label{418}
\|\nabla(U'_n-U'_0)\|^2_{L^2(\Omega)}\leq C\biggl(\|u'_n-u'_0\|^2_{L^4(\Omega)}+\|v'_n-v'_0\|^2_{L^4(\Omega)}\biggr),
\ee
where $C >0$ is a constant.

For (\ref{412}), we have
\be\label{419}
\|\nabla(V'_n-V'_0)\|^2_{L^2(\Omega)}\leq C\biggl(\|u'_n-u'_0\|^2_{L^4(\Omega)}+\|v'_n-v'_0\|^2_{L^4(\Omega)}\biggr).
\ee

From (\ref{418}) and (\ref{419}), we arrive at
\be\label{420}
\|(U'_n-U'_0,V'_n-V'_0)\|_X \leq C\biggl(\|u'_n-u'_0\|^2_{L^4(\Omega)}+\|v'_n-v'_0\|^2_{L^4(\Omega)}\biggr),
\ee
where $C >0$ is a constant. This proves that $(U'_n,V'_n)\rightarrow (U'_0,V'_0)$ strongly in $X$ and the lemma follows.
\end{proof}

We now study the fixed point equation labeled  by a parameter $t$,
\be\label{421}(u'_t,v'_t)=tT(u'_t,v'_t),\quad 0\leq t\leq1.\ee

\begin{lemma}\label{lemm42}
There is a constant $C>0$ independent of $t\in[0,1]$ so that
\be\label{422}
    \|(u'_t,v'_t)\|_X\leq C,\quad 0< t\leq1.
\ee
Consequently, $T$ has a fixed point in $X$.
\end{lemma}

\begin{proof}
When $t>0$, it is straightforward to check that $(u'_t,v'_t)$ satisfies
the equations
\begin{eqnarray}\label{423}
\Delta u'_t &=& \lambda t(\frac{2C_2e^{u'_t}}{\int_\Omega e^{u'_t}dx}-\frac{C_1e^{v'_t}}{\int _\Omega e^{v'_t}dx}-1),\\
\label{424}
\Delta v'_t &=& \lambda t(\frac{-2C_2e^{u'_t}}{\int_\Omega e^{u'_t}dx}+\frac{3C_1e^{v'_t}}{\int _\Omega e^{v'_t}dx}-1)+4\pi t\sum^n_{s=1}\delta_{p_s}(x).
\end{eqnarray}
In the doubly periodic domain $\Omega$, we let $p, q\in\Omega$ so
that
$$u'_t(p)=\max\{u'_t(x)|x\in\Omega\},\quad v'_t(q)=\max\{v'_t(x)|x\in\Omega\}.$$
To facilitate our computation, we adopt the notation \be\label{425}
h'_t(x)=\frac{C_2e^{u'_t}}{\int_\Omega e^{u'_t }dx},\qquad
g'_t(x)=\frac{C_1e^{v'_t}}{\int_\Omega e^{v'_t }dx}. \ee
Then from (\ref{423}), we have
$$0\geq(\Delta u'_t)(p)=\lambda t(2h'_t(p)-g'_t(p)-1).$$
Therefore
$$
2h'_t(p)\leq g'_t(p)+1\leq\frac{C_1e^{v'_t(q)}}{\int_\Omega e^{v'_t}dx}+1= g'_t(q)+1.
$$
Hence, for any $x\in\Omega$, we have
 \be\label{426}
  2h'_t(x)\leq g'_t(q)+1, \quad\forall x\in \Omega.
\ee
From (\ref{424}), using (\ref{426}), we obtain
\be\label{427}
g'_t(q)\leq 1.
\ee
In view of (\ref{426}) and (\ref{427}), for any $x\in\Omega$, we have
\be\label{428}
g'_t(x)\leq 1, \quad h'_t(x)\leq1, \quad x\in \Omega.
\ee
 Set $v'_t=tv_0+w'_t$. Then the
equations (\ref{423}) and (\ref{424}) are modified into
\begin{eqnarray}\label{429}
\Delta u'_t &=& \lambda t(\frac{2C_2e^{u'_t}}{\int_\Omega e^{u'_t}dx}-\frac{C_1e^{tv_0+w'_t}}{\int _\Omega e^{tv_0+w'_t}dx}-1),\\
\label{430}
\Delta w'_t &=& \lambda t(\frac{-2C_2e^{u'_t}}{\int_\Omega e^{u'_t}dx}+\frac{3C_1e^{tv_0+w'_t}}{\int _\Omega e^{tv_0+w'_t}dx}-1)+\frac{4\pi n}{|\Omega|}t,
\end{eqnarray}
where $\Delta v_0=-\frac{\displaystyle4\pi n}{|\Omega|}+4\pi{\displaystyle\sum_{s=1}^n }\delta_{p_s}(x)$. Multiplying
(\ref{429})  and (\ref{430}) by $u'_{t},w'_{t}$ and integrating by parts, respectively, and using (\ref{428}), we have

\bea\label{431}
\|(\nabla u'_t,\nabla w'_t)\|^2_{L^2(\Omega)}
&\leq& \int_\Omega\biggl|\lambda t\biggl(\frac{2C_2e^{u'_t}}{\int_\Omega e^{u'_t}dx}-\frac{C_1e^{tv_0+w'_t}}{\int _\Omega e^{tv_0+w'_t}dx}-1\biggr)\cdot u'_t \biggr|dx\nn\\
&&+\int_\Omega\biggl|\biggl\{\lambda t(\frac{-2C_2e^{u'_t}}{\int_\Omega e^{u'_t}dx}+\frac{3C_1e^{tv_0+w'_t}}{\int_\Omega e^{tv_0+w'_t}dx}-1)+\frac{4 \pi n}{|\Omega|}t\biggr\}\cdot w'_t | dx\nn\\
&\leq&\int_\Omega\lambda (2+1+1)|u'_t|dx+\int_\Omega\biggl((2+3+1)\lambda+\frac{4\pi n}{|\Omega|}\biggr)|w'_t|dx\nn\\
&\leq&\tilde{C}_1\int_\Omega|u'_t|dx+\tilde{C}_2\int_\Omega|w'_t|dx\nn\\
&\leq&C_\varepsilon+\tilde{C}\varepsilon\|(\nabla u'_t,\nabla w'_t)\|^2_{L^2(\Omega)},
\eea
Let $\varepsilon >0$ be small enough, we have
\be\label{432}
\|(u'_t, w'_t)\|_{X}=\|(\nabla u'_t,\nabla w'_t)\|_{L^2(\Omega)}\leq C,
\ee
where $C >0$ is a constant. The existence of a fixed point
is a consequence of Lemma \ref{lemm42}, the apriori estimate
(\ref{422}) and the Leray--Schauder theory. The proof of the lemma
thus follows.
\end{proof}

\section {Further extensions}
\setcounter{equation}{0} \hspace*{1cm}

In this section, we show that our method may be applied to establish
the same existence and uniqueness theorem for multiple vortex
solutions in the $U(3)\times U(3)$ model.

We note that in the study \cite{ABEK} of the non-Abelian multiple vortex
equations (\ref{o2})--(\ref{o5}) the real-valued scalar field
$\kappa$ and the complex-valued scalar field $\phi$ are allowed to
independently generate vortices with their respectively prescribed
zero sets
 \be\label{51}
    Z_\phi =\{p_1,p_2,...,p_n\},\quad Z_\kappa=\{q_1,q_2,...,q_m\}.
\ee
In such a context, we can similarly develop an existence and uniqueness theory for the solutions of the
equations by the same variational methods. To see this, we observe that, with the prescribed zero sets given in (\ref{51}) for the fields $\kappa$ and $\phi$ and in terms of the variables $u=\ln {\kappa}^2$ and $v=\ln|\phi|^2$, the governing system of nonlinear elliptic equations (\ref{o14}) and (\ref{o15}) is modified into
\begin{eqnarray}\label{52}
  \Delta u &=& \lambda (2e^u-e^v-1) +4\pi \sum _{t=1}^m\delta_{q_t}(x),\\\label{53}
  \Delta(u+v)&=& 2\lambda(e^v-1)+4\pi\sum _{t=1}^m\delta_{q_t}(x)+4\pi \ \sum _{s=1}^n\delta_{p_s}(x),
\end{eqnarray}
with the associated boundary condition
\be\label{54}
u,v\rightarrow 0\quad \textmd{as}~ |x|\rightarrow\infty.
\ee
Parallel to Theorem {\ref{theorem031}}, we have
\begin{theorem}\label{theorem51}
The system of nonlinear elliptic equations (\ref{52}) and (\ref{53}) subject to the boundary condition (\ref{54}) has a unique solution for which the boundary condition (\ref{54}) may be achieved exponentially fast.
\end{theorem}
In order to prove Theorem {\ref{theorem51}}, we introduce the background functions as before,
\bea\label{55}
u_0(x)&=&-\sum^m_{t=1}\ln(1+\tau|x-q_t|^{-2}),\\
\label{56}
v_0(x)&=&-\sum^n_{s=1}\ln(1+\tau|x-p_s|^{-2}),\quad\tau>0.
\eea
Then
\bea\label{57}
\Delta u_0&=&-h_1(x)+4\pi\sum^m_{t=1}\delta_{q_t}(x),\\
\label{58}
\Delta v_0&=&-h_2(x)+4\pi\sum^n_{s=1}\delta_{p_s}(x),
\eea
where
\bea\label{59}
h_1(x)&=&4\sum^m_{t=1}\frac{\tau}{(\tau+|x-q_t|^2)^2},\\
\label{510}
h_2(x)&=&4\sum^n_{s=1}\frac{\tau}{(\tau+|x-p_s|^2)^2}.
\eea

We set $u=u_0+w_1,v=v_0+w_2,$ and $f=w_1+w_2$. Then (\ref{52}) and (\ref{53}) become
\bea\label{511}
\Delta w_1&=&\lambda\biggl(2e^{u_0+w_1}-e^{v_0+f-w_1}-1\biggr)+h_1(x),\\
\label{512}
\Delta f&=&2\lambda\biggl(e^{v_0+f-w_1}-1\biggr)+h_1(x)+h_2(x).
\eea

It can be checked that (\ref{511}) and (\ref{512}) are the Eular--Lagrange equations of the action functional
\bea\label{513}
I(w_1,f)&=& \int_{{\mathbb{R}}^2}\biggl\{\frac{1}{2\lambda}|\nabla w_1|^2+\frac{1}{4\lambda}|\nabla f|^2+2\biggl(e^{u_0+w_1}-e^{u_0}\biggr)+\biggl(e^{v_0+f-w_1}-e^{v_0}\biggr)\nn\\
&&+\biggl(\frac{h_1}{\lambda}-1\biggr)w_1
+\biggl(\frac{h_1+h_2}{2\lambda}-1\biggr)f\biggr\}dx.
\eea

It is clear that the functional $I$ is $C^1$ over $W^{1,2}({\mathbb{R}}^2)$ and strictly convex. We can use the methods in \cite{JT} and in the earlier study in the present paper to establish the coercive bounds
\be\label{514}
D I(w_1,f)(w_1,f)\geq C_1\biggl(\|w_1\|_{W^{1,2}({\mathbb{R}}^2)}+\|f\|_{W^{1,2}({\mathbb{R}}^2)}\biggr)-C_2.
\ee
Therefore, it follows that the functional $I$ has a unique critical point in $W^{1,2}({\mathbb{R}}^2)$ which establishes the existence and uniqueness of a classical solution to the system of equations (\ref{52}) and (\ref{53}) subject to the boundary condition (\ref{54}).

We now turn our attention to the existence of multivortex solution over a doubly periodic domain $\Omega$.

Take $u_0$ and $v_0$ over $\Omega$ to satisfy
\be\label{515}
 \Delta u_0=-\frac{4\pi m}{|\Omega|}+4\pi \sum_{t=1}^m\delta_{q_t}(x),
 \quad\Delta v_0=-\frac{4\pi n}{|\Omega|}+4\pi \sum _{s=1}^n\delta_{p_s}(x).
\ee
Then setting $u=u_0+w_1, v=v_0+w_2 $, we see that the equations (\ref{52}) and (\ref{53}) over the doubly periodic domain $\Omega$ become
\bea\label{516}
 \Delta w_1 &=& \lambda(2e^{u_0+w_1}-e^{v_0+w_2}-1)+\frac{4\pi m}{|\Omega|}, \\\label{517}
  \Delta (w_1+w_2) &=& 2\lambda (e^{v_0+w_2}-1)+\frac{4\pi }{|\Omega|}(m+n).
\eea
\begin{theorem}\label{theorem52}  For the vortex equations (\ref{516}) and (\ref{517})
 defined over a doubly periodic domain $\Omega$, there is a
solution if and only if the inequalities
\bea\label{518}
 2\pi (m+n)&<& \lambda|\Omega|,\\
 \label{519}
\pi(3 m+ n)&<& \lambda|\Omega|,
\eea
are satisfied. Moreover, if a solution exists, it must be unique.
\end{theorem}

In the special case when the real scalar field $\kappa$ has no zero,
that is, $m=0$ in (\ref{518}) and (\ref{519}), we recover Theorem
\ref{theorem31}.

To proceed in the formalism of calculus of variations, we use the new variables $g=w _1, f=w_1+w_2$. Then (\ref{516}) and (\ref{517}) take the form
\bea\label{520}
\Delta g &=& \lambda(2e^{u_0+g}-e^{v_0+f-g}-1)+\frac{4\pi m}{|\Omega|} ,\\\label{521}
\Delta f &=& 2\lambda (e^{v_0+f-g}-1)+\frac{4\pi }{|\Omega|}(m+n).
\eea
Integrating these two equations and simplifying the results, we
arrive at the constraints
\bea\label{522}
\int _\Omega e^{v_0+f-g}dx&=&|\Omega|-\frac{2\pi}{\lambda}(m+n)\equiv \alpha_1,\\\label{523}
\int _\Omega e^{u_0+g}&=&\frac{1}{2}(\alpha _1+|\Omega|-\frac{4\pi m}{\lambda})\equiv\alpha_2.
\eea
In order to show that the necessary condition $\alpha_{1}>0$, $\alpha_{2}>0$, which is exactly what stated in (\ref{518}) and (\ref{519}), is also sufficient for the existence of a solution, we recognize that Eqs. (\ref{520}) and (\ref{521}) are the Euler--Lagrange equations of the action functional
\bea\label{524}
I(f,g) &=& \int_\Omega\biggl\{\frac{1}{4\lambda}|\nabla f|^2+\frac{1}{2\lambda}|\nabla g|^2+2e^{u_0+g}+e^{v_0+f-g}\nn\\
 &&+\biggl(\frac{4\pi m}{\lambda|\Omega|}-1\biggr)g+\biggl(\frac{2\pi(m+n)}{\lambda|\Omega|}-1\biggr)f\biggr\}dx.
\eea

Now decompose $f, g$ into $f=f'+\underline{f}, g=g'+\underline{g}$
with $\underline{f}, \underline{g}\in \mathbb{R}$ and $\int_\Omega
f'dx=0,\int_\Omega g'dx=0$. Thus, applying (\ref{522}) and
(\ref{523}), we may rewrite (\ref{524}) in the form

\bea\label{525}
I(f,g)&-&\int_\Omega\{\frac{1}{4\lambda}|\nabla f'|^2+\frac{1}{2\lambda}|\nabla
g'|^2\}dx\nn\\
&=& \alpha_1\ln(\int_\Omega e^{v_0+f'-g'}dx)+2
\alpha_2\ln(\int_\Omega e^{u_0+g'}dx)\nn\\
&&+\alpha_1(1-\ln\alpha_1)+2 \alpha_2(1-\ln\alpha_2).
\eea

It is seen immediately that the right-hand side of (\ref{525}) has a
uniform lower bound in view of the Jensen inequality again. So the
existence of a critical point of (\ref{524}) subject to the
constraints (\ref{522}) and (\ref{523}) follows as before. The
uniqueness of a critical point of (\ref{524}) results from the
convexity of the functional.\\[2mm]



\end{document}